\documentclass{article}
\usepackage[pdftex]{graphicx}
\usepackage{amsfonts}
\usepackage{multirow}
\usepackage{url}

\begin{document}
\title{Cost estimation of a fixed network deployment over an urban territory}
\author{Catherine Gloaguen \and  Elie Cali  }      
\maketitle

\begin{abstract}
Using theoretical results presented in former papers, two prototypes were developed, aiming at estimating within minutes the cost of a fibre network deployment on a given territory. The first one helps defining the limit of an urban territory and computes mathematical parameters representing its street system. The second one gives global information on a fixed network deployment on this territory, namely the probability distributions of distances and attenuation from a node of the network to the final customer, and an evaluation of the deployment cost, once given an architecture and engineering rules. This allows a final user to easily design various network architectures in the tool, to compare different deployment scenarios, and to optimize the budget and the efficiency of the network in a few minutes. The results were compared on two real French urban territories (in Tours and Rouen) to those given by an optimization tool currently used by Orange.
\end{abstract}

\section{Introduction}
Cost estimation is a major issue for operators, especially when their commercial expansion leads them to deploy networks on new territories often using new technologies. Today's world of mobility generates an  amount of challenging problems but does not eliminate the need to consider fixed connections that either constitute whole networks or are part of the infrastructure of mobile ones. High speed  optical connections are nowadays required wherever possible in order to satisfy customers. The problem for an operator lies in the words "wherever possible" that describe the technical feasibility as well as the financial impact of potential solutions; obviously the last one should be minimal. We consider here the design from scratch of a fixed network deployment at urban scale under cost and efficiency constraints.

The description of a realistic telecommunication network deployment involves a huge number of parameters not all equally well documented, spanning fields from geography, regional development, sociology, technology, etc\ldots\, One does not a priori know which of them will play a crucial role for a given analysis such as cost estimation in our case. For example, the physical links of fixed networks (optical fibres or copper wires) are located along the streets and therefore one can assume that it is compulsory to include the underlying road system in the formulation of the problem. But what is its real impact on the sought result? And what precise characteristics of the street system is it important to consider? Exhaustive reconstructions of deployment solutions cannot answer this for computational reasons (precise input data are often missing  and computing time to find optimal solutions is prohibitive) and due to the difficulty to extract the impact of each parameter from a set of solutions. 

Once the architecture and deployment rules are known, optimization methods \cite{Grotschel,Chardy2} compute optimal locations for nodes and optimal ways for paths in order to minimize the cost of a network given the area of deployment. However, for planning purposes on a territory, other methods must be used to determine globally what would be the potentially best architectures and technical choices prior to the above mentioned detailed spatial localization of equipment. One class of such methods uses graph theory and complex networks theory to compute an estimation of the number of nodes and of the trenching length needed for a given urban area and a simple architecture \cite{Maniadakis:streetFTTH}. 
 
 Another class of such methods are macroscopic models based on \textbf{\textit{stochastic geometry}}. Stochastic geometry \cite{Chiu} is the mathematical field of spatial probability and has a wide range of applications \cite{Kendall}. In particular, it allows an instantaneous analysis of modern telecommunication networks, which performances strongly depend on the spatial arrangement of their components, while focusing on global results rather than a detailed description of the network. This approach makes it possible to take into account and to understand the statistical complexity of the system by extracting its structuring parameters and explicitly showing how they impact the statistics of geometry dependent quantities. Here, one is not interested in the precise location of network equipment, but on statistics on quantities that are meaningful for the operator such as distances from customers to equipment, or costs, that can decide on the economic interest and/or technical feasibility of a network design on a particular area. 

The idea of macroscopic models for fixed network is to replace actual locations of objects by random processes that take into account the main principles of their positioning while introducing a variability that represents the effect of numerous and unavoidable local constraints that compel in reality to slightly modify the planned locations. Random processes deal with geometrical objects of any shapes. Their definition explicitly describes some rules and associated parameters to randomly generate realizations of sets of these objects. So called "functionals" (for example the distances between pairs of points) can then be computed on each realization of the set. The aim of stochastic geometry is to derive closed form expressions for the output of the functional, depending on the type and parameters of the processes. In this way, statistics on the geometrical properties of sets of objects can be obtained instantaneously and independently of their number of elements.   

This paper focuses on cost estimation of network at urban scale using the techniques of stochastic geometry. It is organized as follows: Section \ref{sec:methods} recalls the context of existing models developed in Orange Labs and explains the underlying mathematical objects involved; Section \ref{sec:implementation} describes how those objects are organized and integrated in a user-friendly tool to build a faithful macroscopic model of the implementation of a realistic network and its costs; Section \ref{sec:results} provides and discusses results on real cities and Section \ref{sec:conclusion} opens some perspectives. 

\begin{figure}
\centering
\includegraphics{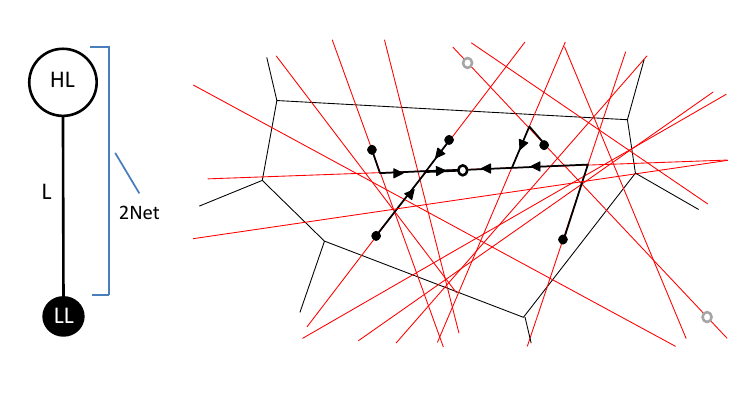}
\caption{(left) Formal architecture of a 2 levels network. The 2Net = \{HL, L\} is structured by LLs but does not include their characteristics in its definition. (right) An example of a network deployed on streets modelled by a PLT. HLs (circles) and LLs (dots) are located on the streets. All LLs located in the serving zone (black polygon) of a HL are connected to it; the physical paths of the links (thick black) follow the streets.}
\label{fig:Fig2Net}
\end{figure}

\section{Macroscopic modelling and methods}\label{sec:methods}
In \cite{Bacelli} a simple cost function for a hierarchical core network which nodes are located in the plane is considered and optimized to deduce the characteristics of the middle layer. To address access networks that describe connections to the customers, the location of networks equipment must be estimated considering the existing road system since the network nodes are located on or close to the streets. To take advantage of the power of stochastic modelling, all the parts of the model including the street system have to be randomized. Subsections below briefly present the work done in the past years concerning the randomization, respectively of the street system, the typical cell, the links and the capacity trees. Discussions and details can be found in the references. 

\subsection{Elementary network} \label{subsec:2Net}
We call \textbf{\textit{node}} a geographical site containing various pieces of equipment, designed to perform specific functions in the network. Nodes of similar functions are referred to by their type such as PB (branching point) or PM (shared access point). This section will be devoted to the model of the elementary two levels network. Its architecture is the simplest one, involving low level nodes (LL), high level nodes (HL) and their formal \textbf{\textit{link}} (L) (Fig. \ref{fig:Fig2Net}%
). LL and HL must be of different types. Engineering rules ensure that any node LL is connected to a unique node HL and specify how the nodes and the physical path of the link are spatially located.  

A \textbf{\textit{2Net}} is the formal representation of a number of HL nodes and their links L deployed on the territory to be served. Note that the characteristics of the links (number, length ...) depend on both HL and LL, but their description does not include any node characteristics. We shall explain below how a 2Net can be modelled by a unique spatial random object, the \textbf{\textit{typical cell}} and its content. Any hierarchical architecture can be represented as an assembly of 2Nets. For example a three levels purely hierarchical architecture from LL to HL is built form (HL, LL\_2, L\_2) and (HL\_1, LL, L\_1) by identification of LL\_2 and HL\_1. The generic aspect of the 2Net notion allows us to conduct generic computations on all network characteristics whatever the architecture (see Section \ref{subsec:network_design}).

\subsection{Models for street systems} \label{subsec:street}
The morphology of an urban street system reflects the history of the city. Ignoring dead-ends, the streets segments of a planar map constitute the edges of a planar tessellation. The cells of the tessellation are generally not uniform: their area, number of edges, etc\dots can vary from one cell to the other. To cope with this spatial variability, and ignoring the finite area of the map, it is viewed as a realization of a random stationary tessellation $\mathcal{T}$ that is fully determined by three independent scalar geometrical quantities averaged per unit area \cite{Moller:tessell}. We chose to represent the morphology of the map by the four-vector $\mathcal{M}_{real}$ composed of quantities than can be easily estimated: the mean number of crossings, of edges, of cells and the mean length of streets. The mean number of edges is included for checking purpose and should equal the mean number of crossings plus cells.

Simple homogeneous random models as Poisson Vorono\"{i}, Poisson Line and Poisson Delaunay tessellation (PVT, PLT and PDT) are good candidates to capture the morphology of a street system. Each one is fully described by a unique scalar intensity parameter $\gamma$. Their corresponding theoretical vector $\mathcal{M}_{theo}^{model}$ can be expressed as a function of $\gamma$, for example: $$\mathcal{M}_{theo}^{PVT}(\gamma)=(2 \gamma,3 \gamma,\gamma,2 \sqrt{\gamma})$$Then, a \textbf{\textit{fitting procedure}} \cite{Gloaguen:fitting} applied to the city map selects the best random model chosen in a set of available models and estimates its parameter by minimizing a distance between $\mathcal{M}_{real}$ and $\mathcal{M}_{theo}^{model}(\gamma)$. Realizations of the selected model look quite different from the real data, but averaged quantities are close. For example: 1855 crossings, 2937 edges, 1083 cells and 439 km streets measured in an area 53.9 km$^2$ of the French city of Pau, correspond to a PVT, $\gamma$=18.2 km$^{-2}$ and theoretical averages: 1961 crossings, 2941 edges, 980 cells and 460 km streets. 

To cope with the frequent non stationarity of cities, a \textbf{\textit{segmentation}} procedure \cite{Courtat:physrev} automatically partitions the city in as many homogeneous parts as desired and fits the best model for each part. Simple Poisson models are sufficient for our final purpose, but better fits can be obtained considering iterated \cite{Maier:iterated}, STIT \cite{Nagel:stit} or Poisson Gabriel \cite{Courtat:these} tessellations. 

\subsection{Nodes locations and typical cell} \label{subsec:typicalcell}
The above defined 2Net will be deployed in an area covered by a street system fitted by a random process. In order to anchor this 2Net to the territory, one has to specify (i) how to locate the nodes with respect to the streets and (ii) how to associate a unique HL to any LL, thus defining the HL serving zone with respect to LLs (Fig. \ref{fig:Fig2Net}). 

The process for nodes is doubly stochastic (Poisson-Cox process) since they are assumed to be randomly positioned following a linear Poisson process on the random process for streets. Processes for HLs and LLs are assumed independent and are described by intensity parameters $\lambda_{HL}$ and $\lambda_{LL}$ under homogeneity assumption (see Appendix). 

Any LL located in the geographical area called the serving zone of the HL will be connected to it. The serving zone of a HL is modelled as its Vorono\"i cell (set of points of the plane closest to this HL than to any other) constructed with the Euclidean metric. Any LL is thus almost surely connected to a single HL. Note that the set of all serving zones is a Poisson-Cox-Vorono\"i-Tessellation (PCVT) of the plane, different from a PVT since the location of HLs is restricted to the streets. We define the \textbf{\textit{typical cell}} as the random object which characteristics obey the statistics of the cells of PCVT. Fast algorithms for simulation of these cells are described in \cite{Gloaguen:typical,Fleischer:typical}. They are based on realizations of the street system supporting the HL and stopping criteria from neighbouring HLs locations. Links and capacity trees discussed below are mathematically defined relatively to those typical cells.

More complicated models using non Euclidean metric for the typical cells and Gibbs models for the location of nodes, could be thought more realistic. But the simpler assumptions above proved to be quite effective and sufficient for our macroscopic approach.

\subsection{Length of links within typical cells} \label{subsec:links}
The links between a HL and all the LLs included in its serving zone are assumed to follow the shortest paths along the streets. Their number as well as their lengths $\ell$ are random. Explicit formulas relating the average $\bar{\ell}$ to the parameters of the streets and nodes random models were proposed in \cite{Gloaguen:networks}. This is not enough for practical purpose on topics such as eligibility (\% of customers closer than a given length threshold to the highest node, ensuring a good quality for communication) that require to know the probability distribution of the length $dis_{LH}(\ell)$.

Closed parametric forms  $dis_{LH}^{model} (\gamma,\lambda_{HL},\ell)$ for the probability distribution of $\ell$ (see Appendix) are derived in \cite{Gloaguen:annales,Voss:densities}. The distribution does not depend on $\lambda_{LL}$ and may be understood as the distribution of lengths viewed from a typical LL, or as the distribution of all the lengths of links L in the network. The formulas include limit values and are valid in the whole range $)0,\infty[$ of parameters. They can be plugged in a software, avoiding lengthy simulations, and adapt to all possible cases.

This abstract construction has been successfully tested in \cite{Gloaguen:spaswin} by comparing simulation results to histograms for lengths of links obtained from real networks. The parametric formulas $dis_{LH}^{model}(\ell)$ provide a quantitative and explicit appreciation of the structuring features of our problem that are the morphology of the street system (represented by the random model) and the nature of the physical path of the link (in straight line or along the streets). 

\subsection{Capacity trees within typical cells} \label{subsec:capacity}
\begin{figure}
\centering
\includegraphics[width=3. in]{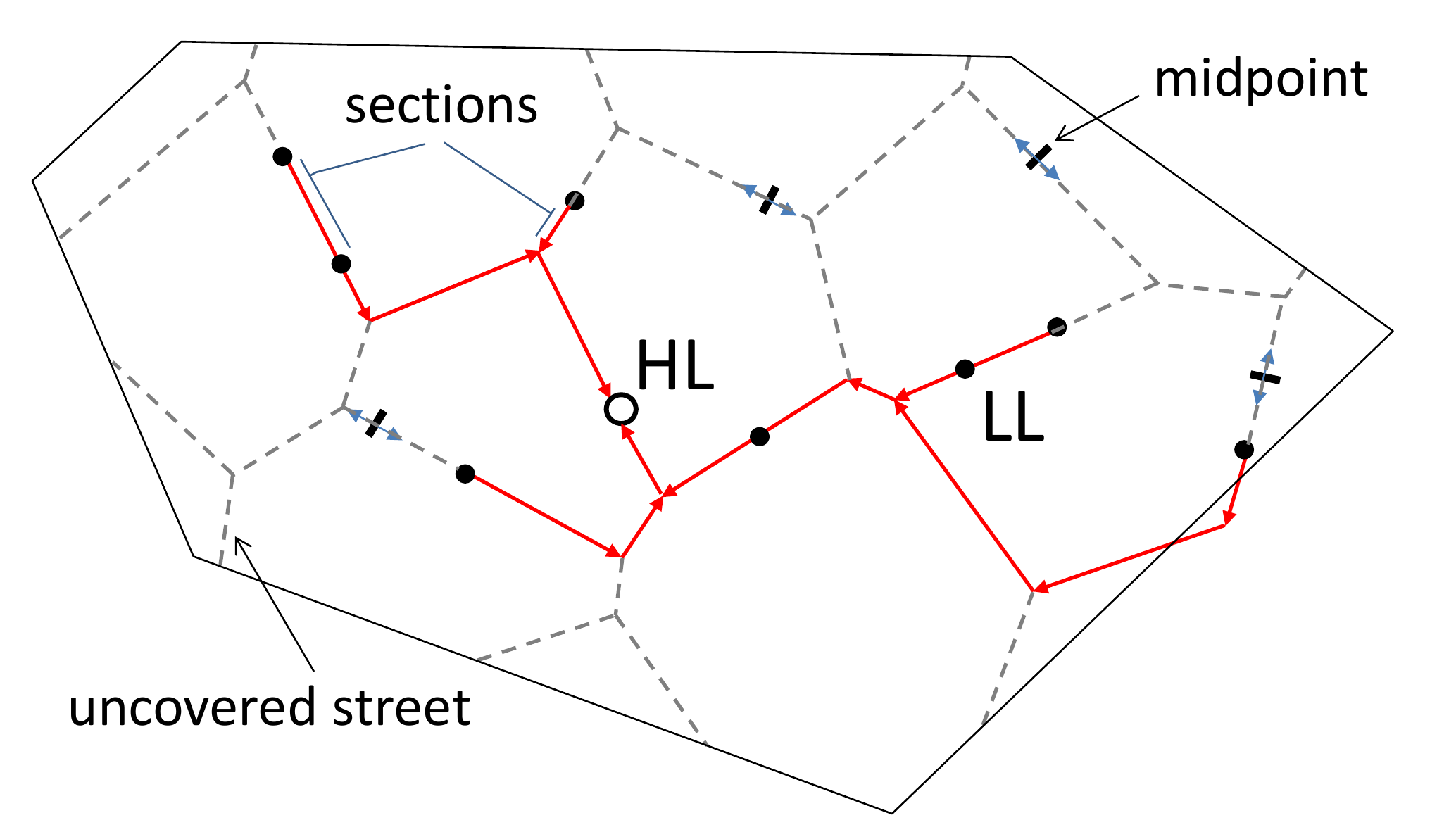} 
\caption{A realisation of a typical capacity tree on a PVT road system (red: path of the link, dotted black: uncovered streets).}
\label{fig:TypicalTree}
\end{figure}

However costs cannot be directly inferred from the distance distributions and one has to consider the multipoint to point tree rooted in HL to all the LLs of its serving zone. This tree has a meaning only if the physical paths of the links follow an underlying street system; then more than one link can share a common street segment. In optical networks, optical fibres are bundled in cables. The cabling policy selects which size of cable(s) will accommodate the \textbf{\textit{capacity}} of each street segment, i.e. the number of fibres running along it. Since the cost of cables depends on their size (for example a 12-fibre cable is quite cheaper than two 6-fibre ones) it is a major issue to study \textbf{\textit{capacity trees}} rooted in HLs. Cabling itself is considered as network engineering and will be addressed in Section \ref{subsec:cable}. In the framework of stochastic modelling, random capacity trees on street systems are extremely difficult to handle. Tentative definitions for structuring characteristics, such as typical sub trees, were investigated in \cite{Voss:capacity} together with estimators for the distribution function of their length. Since an HL is located almost surely on a street segment and not on a crossing, a capacity tree is composed of two half trees that are statistically dependent because they are constrained to live in the same cell. A parametric copula description for their lengths as well as asymptotic results for infinitely sparse or dense networks is studied in \cite{Neuhauser:halftrees}. Sparse random trees with a fixed number of two LLs are fully explored in \cite{Neuhauser:sparsetree}. 

Since we do not have for now generic parametric formulas for the joint probability distributions of the various elements that will enter our cost function (number of cables merging, number of kilometres of cables for each cable capacity), we have to simulate the capacity trees.  The input parameters are $\gamma, \lambda_{HL}$, $\lambda_{LL}$ and the capacity $c$ incoming from each LL, assuming that all LLs bring in a common deterministic capacity. Future work could introduce some variability in this capacity.  

A capacity tree is obtained from a realization of the typical cell and of the LLs (Fig. \ref{fig:TypicalTree}). The root of the tree is the HL. The final leaves are either an edge of the cell or a "midpoint" at which two different paths of equal length to the root HL exist. Note that a path to the HL may need to borrow streets outside the cell. A street element of constant capacity defines a "section" of the tree. Starting from a final leaf, the capacity changes (i) by adding $c$ each time we pass a LL (ii) by adding the capacities of incoming branches at a street crossing.  The vertices of the tree are the street crossings. The set of sections of non zero capacity is the portion of the road system covered by the network. The sections are indexed and each realization of the tree is stored in a text file containing information on vertices (planar location, index of parent vertex), contour of the typical cell, distance of LLs to HL, etc\ldots\, This allows a full reconstruction of the tree and offers the possibility of statistical analysis of all possible single or joined characteristics. For our purpose, only the length of covered streets and the joint distribution (capacity, length of sections) will be used. The limit cases mentioned above are not yet included in the tool that restricts the simulation to a given range of parameters, avoiding cases of too dense or too sparse networks. 

\subsection{Problem formulation}
The sections above introduce typical random objects related to the description of an elementary two levels network. To model the deployment of this network in a realistic setting, one has to precise the geographical area $A$ to be covered (whole city or part of it) and the expected number $Z$ of serving zones (or equivalently of HL nodes) to consider. Added to the model for roads, this characterizes the parameters for the typical cell of averaged area $A/Z$ and the intensity $\lambda_{HL}$ (see Appendix). In real networks, the serving zones in $A$ are not identical and present a variability that is assumed to be correctly represented by the statistics of this typical cell. Rigorously, the covering of $A$ by typical cells should include some constraints on the sum of their areas and border effects should also be considered. For the sought applications, these corrections can be neglected and the very simple approach of combining the characteristics of $Z$ typical cells of identical parameters is sufficient.

To summarize, realistic access networks have their lowest level directly connected to the customers. They can be built by combining elementary 2Nets simply assuming for each of them that the serving zones are Vorono\"{i} areas in which the physical links follow shortest paths on the streets or in the plane. The above mentioned results mainly concern technical and geometrical aspects of the network (nodes, links and capacity trees). 

The theoretical framework being complete, in order to achieve our goal to build an user-friendly tool for cost estimation, one still has to: 
\begin{itemize}
\item enable the user to select a deployment territory, to design an architecture and to specify engineering rules,
\item specify how demography and road data should be selected and processed to be proposed as input for the tool, 
\item enable the user to enter precise formulas allowing to compute the costs,
\item compare the results to data obtained in an operational context and explain the observed discrepancies.
\end{itemize}

\section{Implementation}
\label{sec:implementation}
Two tools were developed with the aim to allow a user to benchmark different architectures and several engineering parameters for each architecture, in order to compare them and to determine the optimal (with respect to costs and eligibility) architecture and engineering rules to deploy in a given city.  

\begin{figure}
\centering
\includegraphics[width=1.8in]{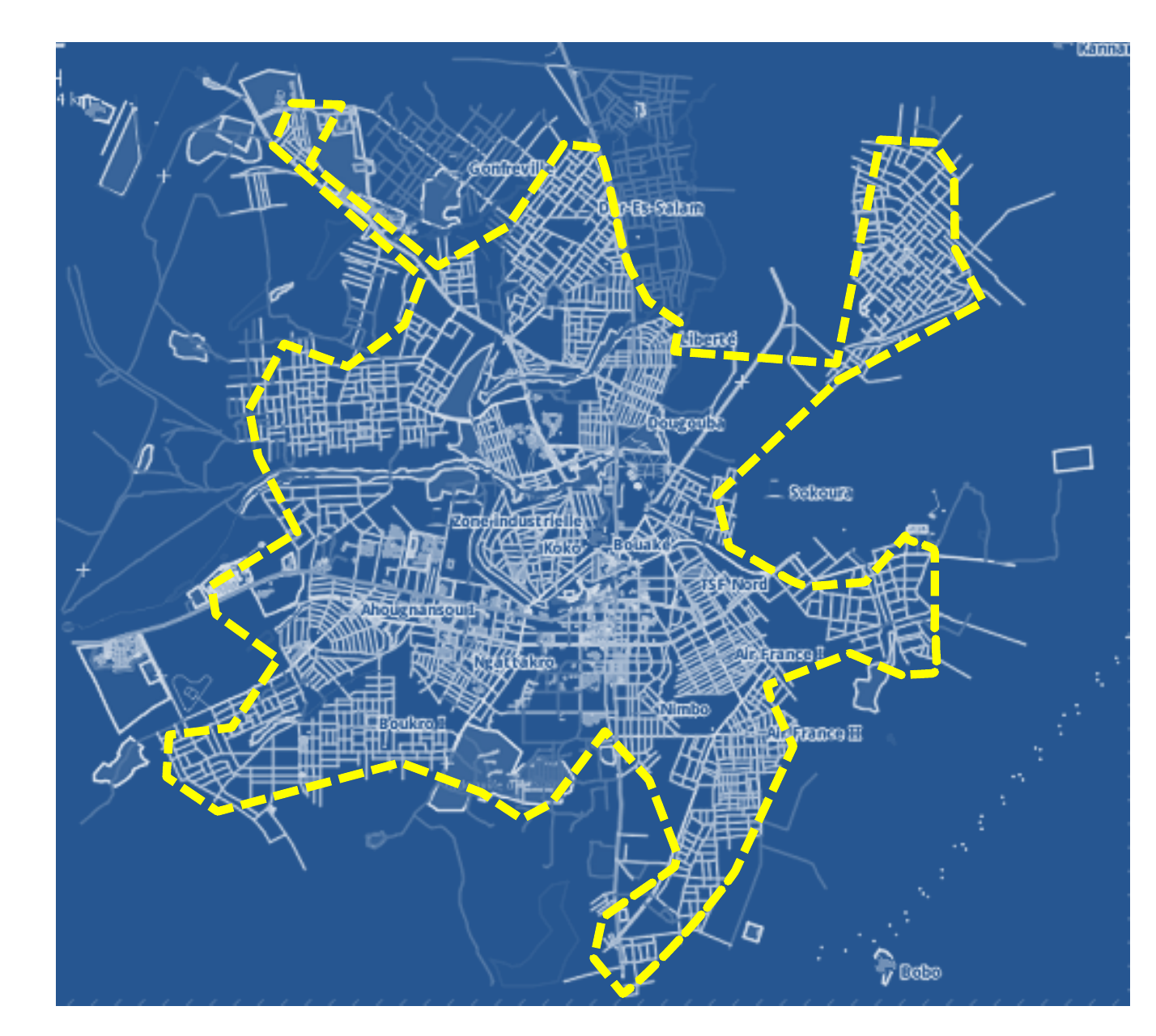}
\caption{Snapshot from OSM Miner showing the street system (grey) for the city of Bouake (Ivory Coast) and the "build-up" area limit computed by the tool (dotted yellow) that can be afterwards tailored by the user. }
\label{fig:osmminer}
\end{figure}

\begin{figure}
\centering
\includegraphics[width=2.5in]{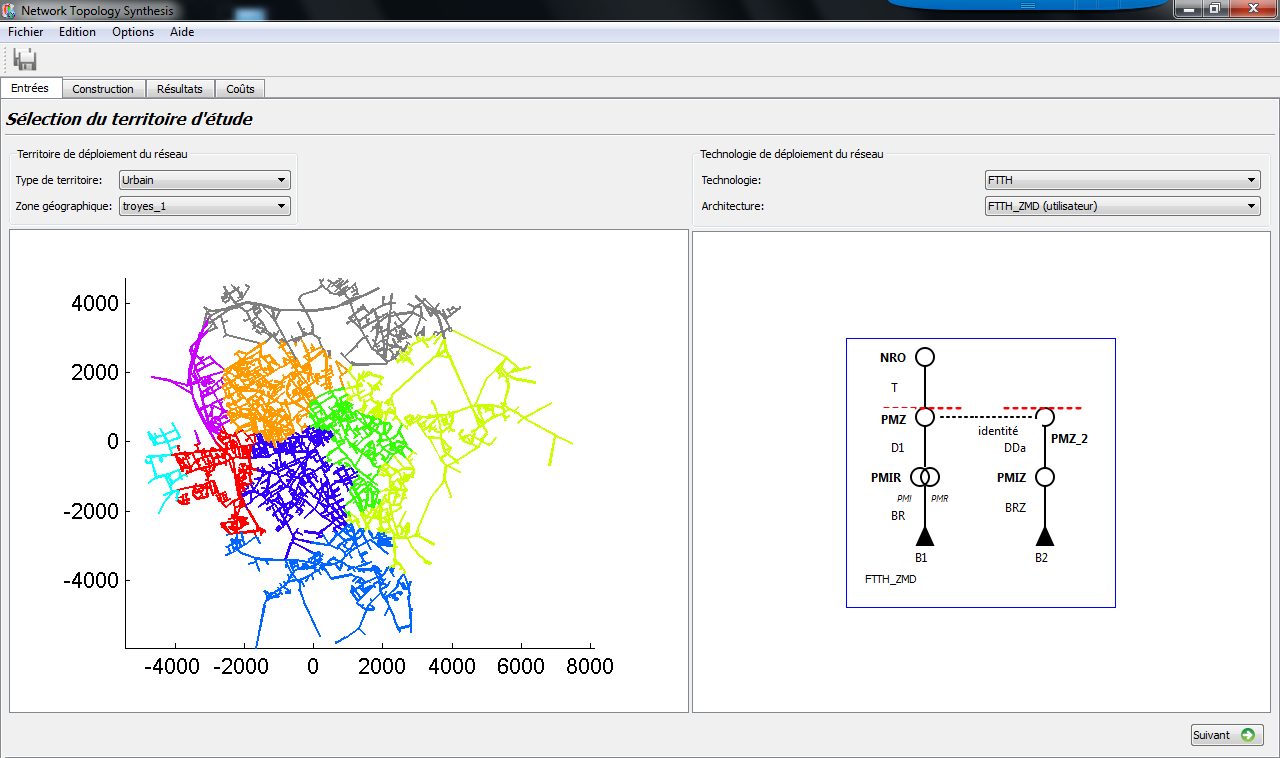}
\caption{NTSTool prototype: the user can choose his territory (imported from the OSM Miner prototype), choose his architecture or design a new one, choose a scenario or enter a new one, and compute the probability distribution of the length of the links and the deployment costs}
\label{fig:ntstool}
\end{figure}

The \textbf{\textit{OSMMiner}} prototype (Fig. \ref{fig:osmminer}) is used first to acquire the territory data, choose the best model to fit them and compute the parameters of the model (Section \ref{subsec:territory}). Then, proceeding with the \textbf{\textit{NTSTool}} prototype, the user selects a territory and designs or selects an architecture (Fig. \ref{fig:ntstool}); the tool then implements the methodology described above to compute the distance and attenuation distributions and to estimate the deployment costs. NTSTool is designed to be generic for urban areas and can potentially address a variety of situations. Sections \ref{subsec:network_design} to \ref{subsec:costs} below focus on optical access networks.  

\subsection{Territory data acquisition and formatting}
\label{subsec:territory}
The road system of the territory is extracted from the Open Street Map (OSM) database \cite{OSM} since it has a worldwide coverage with a very detailed description of the various types of roads. Its rich typology allows to select the types of streets that will support the network wires while suppressing the useless or duplicated ones (eg. cycle ways or paths). Moreover, OSM comes with several free applications and is free of charge. OSM also provides information about buildings, but they are not always complete nor accurate.

The administrative boundary of a given municipality generally differs from the  "build-up" area limit inside which the streets are more or less continuously lined with buildings. The "build-up" area sometimes includes part of the neighbouring municipalities and contains most of the population of the city. It is thus the area of interest from the point of view of network deployment. A java code has been developed to automatically determine this limit using information on buildings. Since there are regions where no information on the buildings is available, the limit can also be deduced from the road system information only. The method is to sum up the number of buildings or streets in each square of a regular grid of adjustable size and to first eliminate the squares with a sum less than a threshold given by the user, and second the isolated squares (those with no more than one neighbour). A concave hull of the set of remaining squares is determined using \cite{Concave} and an algorithm directly inspired by \cite{Duckham}. The limit (Fig. \ref{fig:osmminer}) is given in the form of a standard OSM  ".poly" file  which can then be tailored if necessary by free tools like JOSM \cite{JOSM}. 

Finally, to recover a planar graph for the town, i.e. the streets inside the build-up area, the longitude-latitude coordinates given by OSM are projected on the tangent plane at the center of the territory (which is small enough to neglect distortions). Following the methodology described in Section \ref{subsec:street}, dead-ends are removed in order to get a tessellation \cite{Courtat:these}, the town is segmented in as many parts as desired, and the best model for each part is determined with its intensity parameter (for now, one can chose among three tessellation models: PVT, PDT and PLT). The results are stored in a "\textbf{\textit{voirie}}" file constituted by as many lines as there are parts of the town, plus one line representing the town as a whole (Table \ref{tab:voirie}).

\begin{table}
\renewcommand{\arraystretch}{1.3}
\caption{Example of a "voirie" file for the French city of Nantes. Part1 is the dense core of the city.  }
\label{tab:voirie}
\centering
\begin{tabular}{cccc}
\hline 
part & area (km $^2$) & model & intensity (km $^{-2}$)\\ 
\hline 
1 &  2.181 & PVT & 125.642 \\ 
2 &  10.789 & PVT & 38.101 \\ 
3  & 40.011 & PVT & 13.877 \\ 
total & 52.981 &  PVT & 23.142  \\ 
\hline 
\end{tabular} 
\end{table}

In the same way, information about the households on the territory are stored, if available, in an "\textbf{\textit{habitat}}" file. For French cities, a complete description of the number of households contained in every building in all parts of the town can be obtained from various sources. But for countries where no such information is available, the user will be asked by NTSTool to provide its estimate of the number of households in each part of the town.

\subsection{Designing the network}
\label{subsec:network_design}

\subsubsection{Architecture}
An architecture designer is included in NTSTool , which enables the user to build his own architecture and immediately use the tool on this new architecture. 

\begin{figure}
\centering
\includegraphics[width=2.5in]{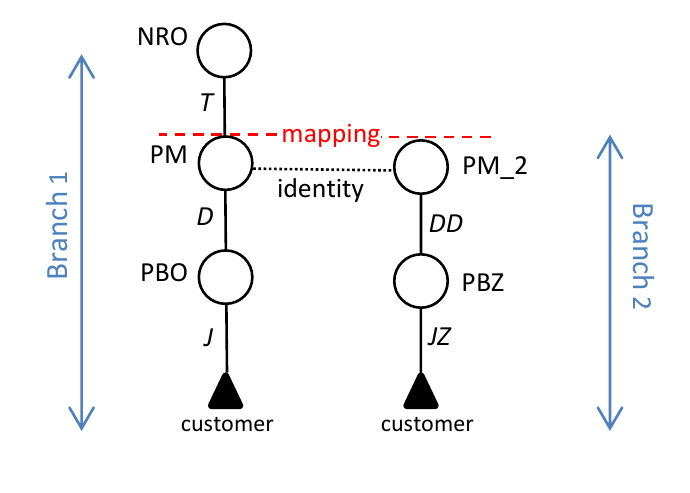}        
\caption{The ZMD architecture: there are two branches, five 2Nets: (NRO,T), (PM,D), (PBO,J), (PM\_2,DD), (PBZ,JZ), a mapping line above the PM node (the (NRO,T) 2Net should be considered on the whole town while the other 2Nets should be considered separately on each part of the town), and an identity connector (PM and PM\_2 are the same nodes, wires of branch 1 go through the PM node to the NRO, while wires of branch 2 pass through the PM site and through the PM node).}
\label{fig:archi}
\end{figure}

A general network architecture may have several branches (Fig. \ref{fig:archi}). Each branch will be represented as an independent set of 2Nets, the highest type of node of a secondary branch being identified with a type of node of the primary branch by an operation called \textbf{\textit{connector}}. A connector can be called either \textbf{\textit{identity connector}}, which means that the wires from the secondary branch will physically pass at the location and through the equipment of the common node, or \textbf{\textit{passage connector}}, which means the wires will physically pass at the location of the node without passing through the equipment (in which case, there is no signal attenuation at this node for the secondary branch). 

Finally, a \textbf{\textit{mapping}} line is defined on the architecture. The 2Nets above the line are considered at the scale of the whole town (using the parameters of the last line of "voirie" file) and the 2Nets below are considered on each part of the town as listed in the "voirie" file described above.

\subsubsection{Engineering}
Orange chose to deploy optical fibre with a "Passive Optical Network" (PON) architecture. A "1:$n$ splitter" can be placed in each node to divide the signal from one single fibre to $n$ ($2, 4, 8, 16$ or $32$) fibres. Since each customer is connected to a fibre in the lower 2Nets, this guarantees a reasonable number of fibres at each level of the network. In order to deploy this network architecture through the town, this and additional information below must be provided by the user in a so-called \textbf{\textit{scenario}}. 

One needs first to describe the general engineering rules that will decide on the total number of nodes in each part of the town. According to the rules chosen, the number of nodes of a given type can then be either directly entered by the user or deduced from the other data (the number of households to be served, or the number of nodes of the inferior level, or the number of wires coming out of the node, etc\ldots). Then, in order to complete the geometrical description of 2Nets, one has to specify how to compute the length of each link. A link does not necessarily follow the streets: for instance, the link from the last node to the customer is generally (all or part  of it) inside a building. So, when defining a link, the user is offered a choice between several possibilities: as the crow flies, following the streets, with constant distance or with randomized (following a normal distribution) distance.

The scenario also specifies the linear attenuation of the fibre (dB/km) and all possible causes that can generate a signal attenuation when passing through each type of node. 

\subsection{Distances and attenuation}
\label{subsec:distances}
A 2Net is associated to a typical cell and thus to the probability distribution of the length of its link. If two 2Nets are one on top of the other in the architecture, the probability distribution of the length of the resulting link will be the convolution of the two probability distributions. The distribution of the total length of a branch is obtained by successive convolutions. The distribution of two branches in parallel is the mean of their distributions pondered by the number of customers. The 2Net notion being generic, this algorithm can be applied to any architecture designed by the user.

\begin{figure}
\centering
\includegraphics[width=3.0in]{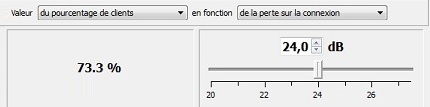}    
\caption{Example of an NTSTool output: the user can choose an attenuation threshold and see the percentage of customers under this threshold in the town}
\label{fig:pertes}
\end{figure}
From this point, the distribution of linear attenuation along the fibres can easily be computed; there remains to compute the attenuation at nodes. A splitting in $n$ fibres in a node corresponds to a signal attenuation of $10 log(1/n)$ dB.  By construction, a node in NTS contains various pieces of equipment that have to be connected, spliced, etc\ldots\, These operations are never perfect nor identically performed and can also generate some signal attenuation. Each type of attenuation, including splitting, is modelled in NTS by a random variable which law (Uniform or Normal) and parameters can be specified. The probability distribution of the total attenuation on each branch is the convolution of all the attenuation distributions associated to nodes and links.  

Thus, given an architecture and a scenario, we are able to compute the attenuation distribution for the whole network, i.e. to say that $x$ percent of the customers in the city will be connected to the highest node with a signal attenuation less than $y$ (dB) to be chosen by the user (Fig. \ref{fig:pertes}).

Note that the convolutions are computed using a Fast Fourier Transform algorithm. The distribution functions have to be sampled with a sufficient number of $2^n$ points, so that the maximum value of the cumulative function of the final result is close to unity. The desired precision (ex. 0.999) is entered by the user. If necessary, the sampling is automatically refined ($n \rightarrow n+1$) to guarantee this goal. The detailed results (including all the sample points of the probability distribution curves) are stored as Excel files for individual 2Nets, branches and the whole network for further analysis. 

\subsection{Cabling}\label{subsec:cable}

\subsubsection{Cabling policy}\label{subsec:policy}A realization of the capacity tree (see Section. \ref{subsec:capacity}) specifies the number of fibres running along each section of the street system covered by a 2Net. Optical fibres are gathered in cables of standard n-fibre types. The cabling policy decides which type(s) of cables to install to accommodate all those capacities. 

Cabling a network is an extremely difficult problem \cite{Angilella:cablage}, even assuming as in NTS that each 2Net can be cabled independently. Note that this assumption is made in most optimization tools and its impact on final cost is judged small by operational teams.  Rigorously speaking, cabling policies should tend to minimize the global cost of cables that depend directly or indirectly on their length, number and types. This means that solutions favouring long big cables even with unused fibres could be better than solutions with shorter and cheaper cables (that generate supplementary costs when merged). NTS prototype uses the simplest policy~: a section of capacity $c$ is cabled with a single n-fibre cable with the lowest $n \ge c$. It can happen that sections close to the HL have a capacity greater that the highest available type of cables. In that case, an additional cable is added to the highest capacity one. 

\subsubsection{Simulations}\label{subsec:simulations}
Now, capacity trees have to be simulated. Some constraints on the parameters ($\gamma, \lambda_{HL}, \lambda_{LL}$) ensure that the simulation time is not prohibitively long (most architectures in urban area are compatible with these constraints). The final user is required to enter a value for the number $N$ of simulations for each 2Net. As a hint, the tool gives him the choice $N_{a}$ that will insure at $95\%$ that the error on the mean values of the tree characteristics (mainly covered road system and length of segment capacities) will be less than a chosen percentage $a$ (following the central limit theorem), and the estimated time of simulation for the value he chooses. Of course, the error on the final result will also depend on the standard deviation of the distribution of each parameter, but this is limited because the sum of the surfaces of the real Vorono\"i cells is equal to the number of cells times the surface of the typical cell. Using the prototype, we could observe that very low values for $N$ ($N=5$ or $10$) already give very good results.

The joint distribution of (types, length) of cables for a 2Net is thus easily computed from the joint distribution of (capacity, length of sections) recovered from all the $N$ simulations. The required time to compute the cabling system for a whole 3-levels network does not exceed ten minutes.

\subsection{Cost computation}
\label{subsec:costs}

\subsubsection{Generic form}
To get the total deployment cost, we need to determine the cost function. This is mainly a question of expertise and know-how of the operator. However, any cost function can be described as a sum of basic bricks:
\begin{center}
 WU * data\_NTS * coeff 
\end{center}
where: 
\begin{itemize}
\item[1)] WU is a Work Unit, i.e. an unitary cost for an article, expressed in a given currency by the final user of the tool (ex: cost of a PMZ in Euros, cost of workforce for laying one kilometre of a 6-fibre cable...),
\item[2)] data\_NTS is a data, already available in NTSTool at this stage, from which the number of elements of the WU can be recovered.(ex: number of households, number of nodes of a given type, etc\ldots),
\item[3)] coeff is a coefficient given by the user (ex: margin multiplier).  
\end{itemize}

A real cost function is usually  composed of a huge number of such basic bricks (one hundred is a rough estimate). The NTSTool prototype includes a "cost function builder" associated to the architecture that allows the user to define his own cost function. In order to ease his task as well as to give explicit and detailed results, it seems sensible to split the costs in four \textbf{\textit{categories}}: material, workforce, civil work (streets trenches etc\ldots) and studies (paperwork); this for as many levels as there are 2Nets in the architecture.  Basic bricks allow to consider a variety of costs as for example:
\begin{itemize}
\item the study costs for aerial Distribution. WU is the elementary study cost for aerial distribution in Euros; data\_NTS is the number of households and coeff is the \% of aerial distribution  (the distribution can also be buried or on fa\c{c}ades), 
\item the civil work costs for chambers. Chambers are manholes used to install new cables or monitor the bundles of cables running under the pavement. They have to be explored prior to any deployment in order to check the space for cables, and thus generate costs. WU is the civil work cost for one chamber in Euros; data\_NTS is the length of covered streets and coeff is one over the mean interval between chambers.
\end{itemize}

Note that the cost of cables writes in a special way. For each link, a family of available cables and associated costs (material and workforce) for one meter of each type of cables, are provided by the user. The cabling procedure (Section \ref{subsec:cable}) computes an histogram of (types, length) of cables for each 2Net that is easily converted in costs.  

Another splitting of costs results in fixed and variable parts may be useful to evaluate the impact of architecture and/or scenario choices on the global cost. Variable costs depend on the architecture and fixed costs - studies and junction 2Nets - depend only on fixed parameters like the number of households to serve in the city. 

\section{Results} \label{sec:results}
\subsection{Comparison setting}
\textbf{\textit{GPON Optimizer}} \cite{Chardy2} is an Orange optimization tool that determines the optimal locations of the nodes, the exact paths of the fibres and the optimal cabling scheme on the real cities' streets given the architecture and the engineering constraints in order to minimize the deployment global cost. GPON exclusively considers the architecture required by the French regulator Arcep \cite{Arcep} for deployment in "less dense areas" of France (Fig. \ref{fig:archi}) with the standard deployment rules used by Orange for fibre deployment (mainly 1:32 splitter in the PM and 1:2 splitter in the NRO (central office), and a mean of 340 households for a PM). An additional regulator rule demands operators to reserve two extra-fibres for competitors for each fibre going from the PM to the NRO, so that the transport cables out of the PM are always 36-fibre ones. The optimization process being complex and thus lengthy, GPON is limited to rather small areas. 

\textbf{\textit{NTSTool}} is conceived to provide statistical results representative of the deployment of a network on a sufficiently large and/or with sufficient spatial variability territory. The above architecture and scenario are easily described for NTSTool, including the additional regulator rule. 

\begin{figure}
\centering
\includegraphics[height=3. cm]{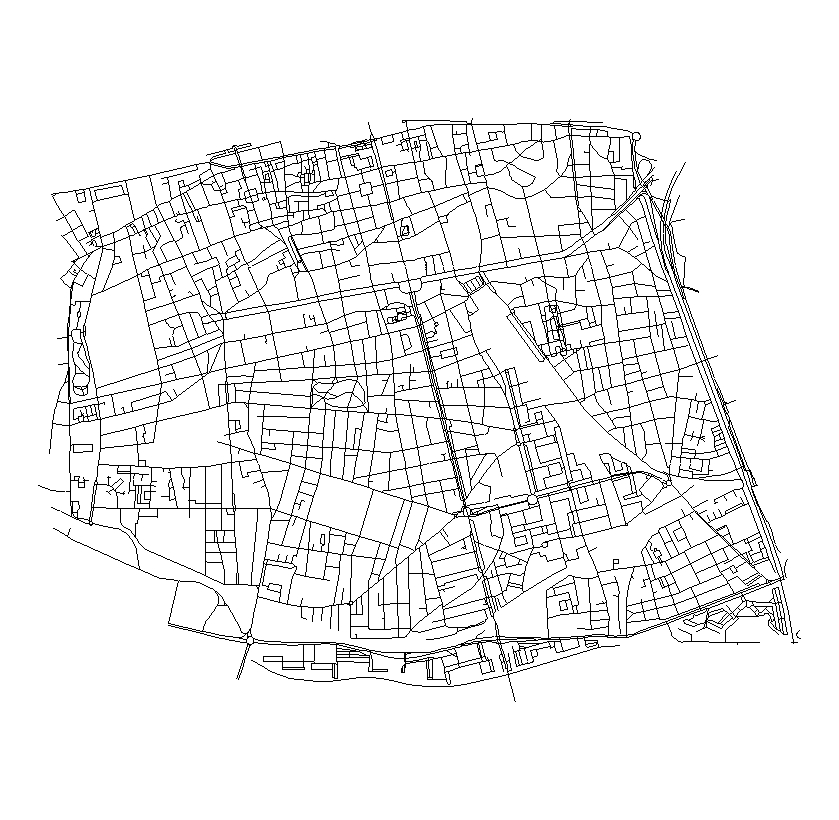} \includegraphics[height=3.5 cm]{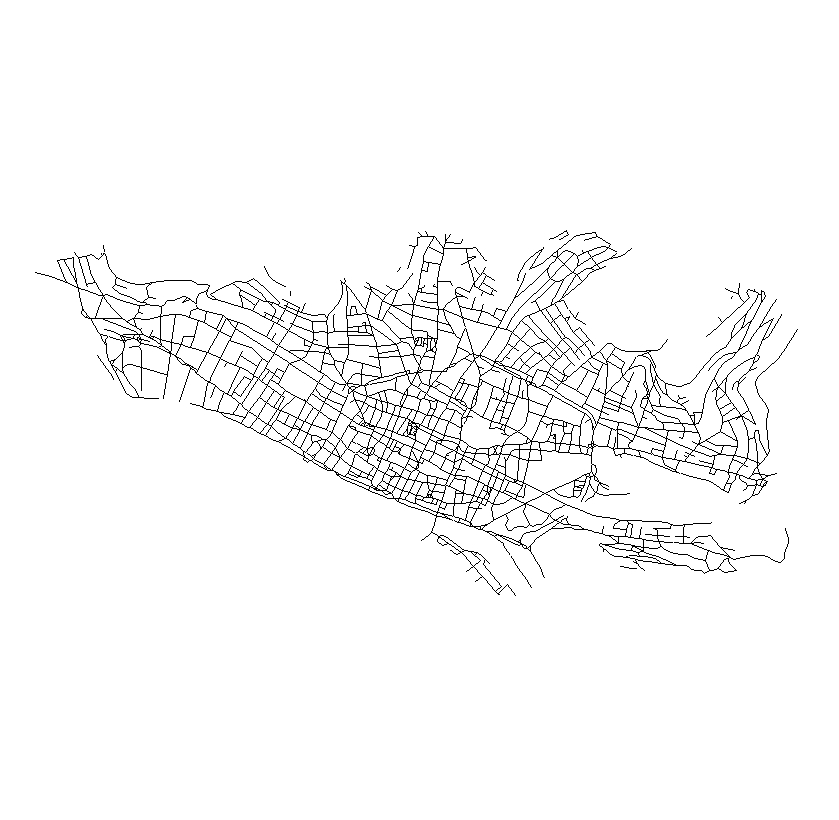}
\caption{Deployment territories used for comparing NTS and GPON. (left) 185 km streets on 9.2 km$^2$ extracted from Tours. (right) 199 km streets on 8.6 km$^2$ extracted from Rouen.}
\label{fig:mapToursRouen}
\end{figure}

In order to check the plausibility of our results, we compare them with the results obtained by GPON Optimizer on parts of two french cities Tours and Rouen. The tests are conducted on one NRO serving zone (Fig. \ref{fig:mapToursRouen}) that constitutes a standard area for GPON Optimizer use, but is at the limit range of size for NTS.  Note that the secondary branch of the architecture does not exist in the territories under study since they do not contain sufficiently huge buildings and that due to their small size they are not segmented (one line in the voirie file). The NTS cost function is built with the same WU than those given in GPON Optimizer.

The results are given for Tours and Rouen, respectively on transport (T), distribution (D) and junction (J) 2Nets. The time needed for 3500 simulations on the D links and 300 simulations on the T link with NTSTool was a few minutes for Tours and for Rouen (while the time needed for GPON Optimizer tool is of course much larger: several scores of hours). 

\subsection{Special case of the last link}\label{subsec:lastlink}
On each branch of the architecture, the last link, between the lowest level node and the final customer, is composed of two parts: one inside the building and one between the building and the NTS node located on the street. In France, the lowest level node is generally placed inside the buildings, while NTS locates all the nodes on the streets, which means that while the first part should be included in the J link, the second one is part of the D link.

The first part is generally a standard kit of fixed price for each household and the workforce to install it is forfeited. Its cost depends only on the number of households to be  equipped in the city and is easily computed. The second part is a cable that can contain a variable number of fibres to accommodate the number of households in the building (one fibre for each household) and runs between the building and the street (and more precisely the closest chamber in the street). A random variable length could have been associated to this type of link, but a fixed length of 15 meters estimated in french cities as a mean proved to be enough. Two cabling policies were considered, compatible with FTTH deployment~: using only 12-fibre cables (policy 1) or using cables adapted to the number of household in each building (policy 2). This has of course some influence on the cabling cost. Policy 1 gives a rougher estimate but may be used when precise information on buildings is unknown (for instance in some African countries).

\subsection{Causes of differences on the cost evaluation}
The NTS cost function is built with the same WUs than those given in GPON Optimizer, but there are still well identified major causes that stand for the differences between NTS and GPON results in the test.  

\textbf{\textit{Errors NTS vs reality}}. Cause \textbf{($C_1$)} comes from WUs that cannot be represented at all or can only be estimated. For example, there exist some network pieces of equipment such as "PA" (Access Point not containing any network equipment) for which it is impossible to write a rule with the scenario builder of NTS. All such WUs that cannot be represented in NTS are gathered and referred to as "Uncounted". It happens also that some WUs have to be approximated since they strongly depend on unknown minute details of the deployment. For instance, the civil works cost of chambers mentioned above depends on the exact number of chambers to be inspected and on their depth. This data being unknown, the number of chambers is estimated assuming a mean 100 meters distance between chambers and taking the cost of inspection for small depths.
\begin{figure}
\centering
\includegraphics[height=3.3 cm]{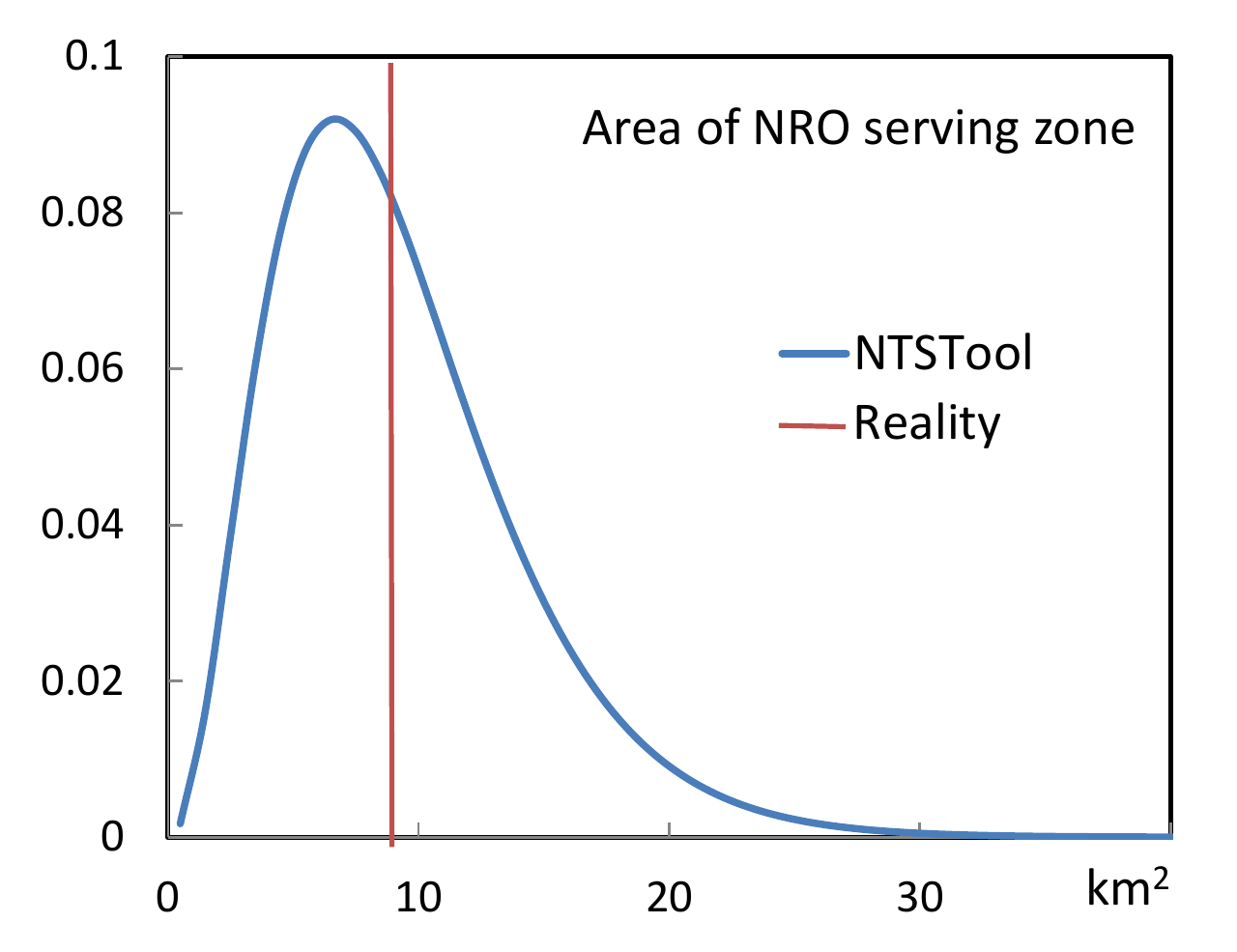} \includegraphics[height=3.3 cm]{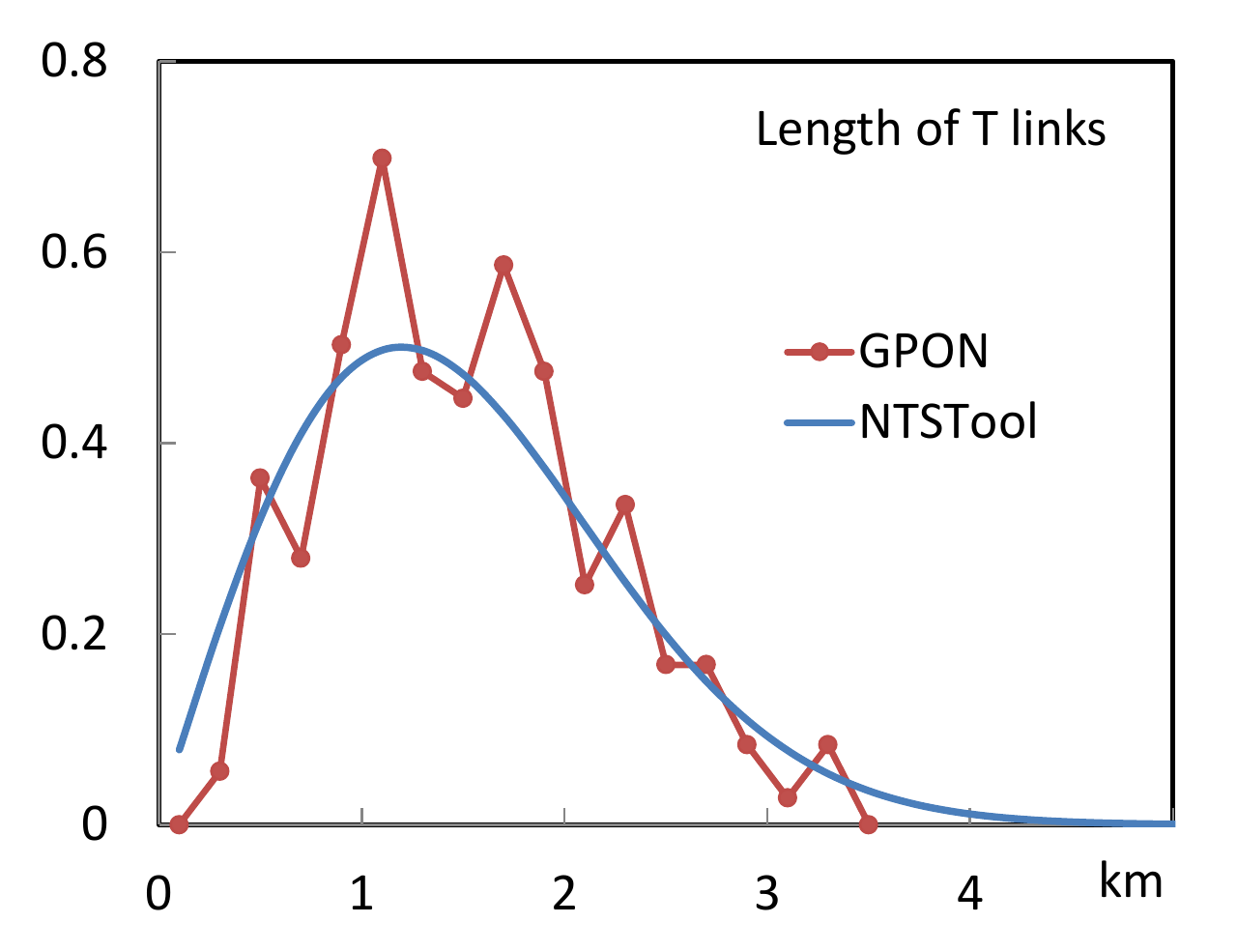}
\caption{Transport 2Net in Tours. Probability distributions of the area of the NRO serving zone (top) and of the length of T Links (bottom).}
\label{fig:randomTours}
\end{figure}

Cause  \textbf{($C_2$)} is due to NTS methodology itself, based on randomness. NTS proposes probability distributions for relevant quantities that can compare to reality only when the "law of big numbers" holds. This statistical nature of NTS results is illustrated by the geometrical characteristics of the Transport \{NRO, T\} 2Net. Recall that the territory area is in reality one NRO serving zone of deterministic area $A_T$. It is modelled in NTS by a random typical cell having the same average area $A_T$ and the probability distribution of Fig. 
\ref{fig:randomTours}. In Tours, the averaged number of LLs connected to one NRO is more than a hundred. Then even considering a single NRO cell, the probability distribution of the lengths of T links provided by NTS (Fig. \ref{fig:randomTours}) is close to the one obtained by GPON that optimizes the locations of the LLs.

\textbf{\textit{Differences NTS vs GPON}}. GPON Optimizer uses a much more elaborated cabling policy than NTSTool's one described in Section \ref{subsec:cable}. This is the cause \textbf{($C_3$)} of some differences that will probably remain even if more sophisticated policies are  integrated in the tool. Cause \textbf{($C_4$)} is produced by some heuristic cabling strategies used by GPON in order to reduce the computing time. The final class of causes \textbf{($C_5$)} is that GPON itself is also a representation of the reality that introduces its own assumptions and depends on detailed databases not always complete nor accurate. 

Given a 2Net and a cost category, the difference is computed as $\bigtriangleup$=(NTS-GPON)/GPON, with positive values whenever NTS costs are greater than GPON ones. For "Uncounted", $\bigtriangleup$ is computed as minus the ratio of GPON costs for Uncounted divided by the total GPON cost.

\subsection{Discussion on Tours }
This section analyses the results for T, D and J 2Nets in Tours and explains the major observed differences with respect to the causes introduced above. 

\begin{figure}
\centering
\includegraphics{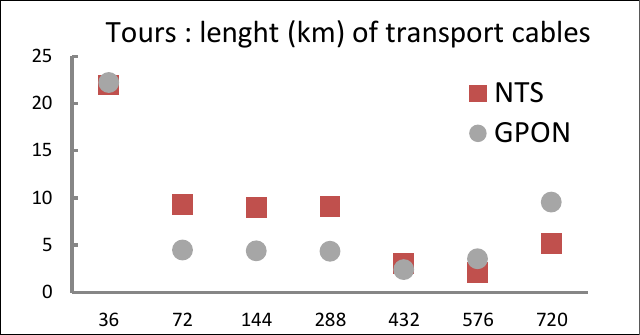}
\caption{The histogram of transport cables in Tours: the x-axis gives the various types of cables used for transport.}
\label{fig:cables_trans_Tours}
\end{figure}

The \textbf{\textit{Transport}} cables histograms for NTSTool and GPON (Fig. \ref{fig:cables_trans_Tours}) shows that NTS underestimates the 720-fibre cables and correspondingly overestimates the 72, 144 and 288-fibre cables. This is mainly explained by \textbf{($C_3$)} because GPON considers over-sized 720-fibres cables if this eventually proves cheaper. 

\begin{table}
\renewcommand{\arraystretch}{1.3}
\caption{Variable costs in Tours, detailed difference $\Delta$ for transport} 
\label{tab:trans_Tours}
\centering
\begin{tabular}{ccc}
\hline
Category & & $\Delta$ \\
\hline
\multirow{3}{*}{Material} & Nodes &  -2\% \\
& Cables & -8\%\\
& Uncounted & -4\%\\
\hline
\multirow{3}{*}{Workforce} & Nodes & 2\%\\
& Cables & 17\%\\
& Uncounted & -41\%\\
\hline
Civil work & Chambers & -29\%\\
\hline
\hline
Total variable&  & -21\%\\
\hline
\end{tabular}
\end{table}
Regarding the variable transport costs (detailed in Table \ref{tab:trans_Tours}), note that the  "Uncounted" lines \textbf{($C_1$)} are by far the main source of differences. The covered part of the road system -and thus the cable workforce- is overestimated in NTS \textbf{($C_2$)}. The total number of chambers is slightly underestimated and they are all associated to small depths \textbf{($C_1$)}; this explains the underestimation of the civil works cost. However, the civil work costs being quite low compared to the other costs, the impact on the total cost is weak. Table \ref{tab:Tours} synthesizes the differences for the fixed (studies), variable parts and total T cost and separately for the cable costs (material and workforce). Overall, for transport in Tours, the cable costs are only 2\% lower for NTS, while the variable costs are much underestimated due to the "Uncounted". 

\begin{table}
\renewcommand{\arraystretch}{1.3}
\caption{Total costs in Tours: difference $\bigtriangleup$ between NTS and GPON. Policy 2 is used for Distribution.}
\label{tab:Tours}
\centering
\begin{tabular}{cccc||c}
\hline 
 & Fixed & Variable & Total & Cables \\ 
\hline 
Transport & 0 \% & -21 \% & -8 \% & -2 \% \\ 
Distribution &0 \% & -8 \% & -7 \% & -8 \% \\ 
Junction &-1 \% & 0 \% & -1 \% & 0 \%\\ 
Total &-1 \% & -9 \% & -4 \% & -6 \%\\ 
\hline 
\end{tabular} 
\end{table}

\begin{figure}
\centering
\includegraphics{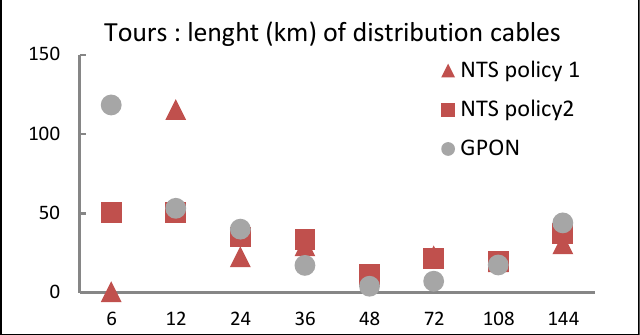}
\caption{The histogram of distribution cables for Tours. NTS adds 15 m of cables from street to buildings: 12-fibre (policy 1) or n-fibre depending on household number (policy 2). The x-axis gives the various types of cables used for distribution.}
\label{fig:cables_dist_Tours}
\end{figure}

The \textbf{\textit{Distribution}} cables histograms (Fig. \ref{fig:cables_dist_Tours}) compares GPON to NTS and shows the impact of the cabling policy for the last link (Section \ref{subsec:lastlink}). GPON overestimates the number of 6-fibre cables \textbf{($C_5$)}. NTS predicts no 6-fibre cables and a big number of 12-fibre cables in policy 1. With policy 2 NTS is close to GPON apart from the 6-fibre cables. The main differences come again from the "Uncounted" \textbf{($C_1$)} which are fewer for distribution and from the civil works costs (same reason than for T). Here however, the covered part of the road system is underestimated in NTS (and with it the cable workforce), due to the fact that the dead-ends are not accounted for \textbf{($C_1$)}. This can be partly corrected by taking into account in the tool their total length that can be saved while analyzing the city prior to the fitting step. 

\begin{table}
\renewcommand{\arraystretch}{1.3}
\caption{Variable costs in Tours, detailed difference $\bigtriangleup$ for distribution, policy 2}
\label{tab:dist_Tours}
\centering
\begin{tabular}{ccc}
\hline
Category & &  $\bigtriangleup$  \\
\hline
\multirow{3}{*}{Material} & Nodes &  1\% \\
& Cables & 0\%\\
& Uncounted & -8\%\\
\hline
\multirow{3}{*}{Workforce} & Nodes & -1\%\\
& Cables & -14\%\\
& Uncounted & -10\%\\
\hline
Civil work & Chambers & -17\%\\
\hline \hline
Total variable& & -8\%\\
\hline
\end{tabular}
\end{table}

The \textbf{\textit{Junction}} 2Net generates fixed costs deduced from the number of households. The 2\% difference comes from some Uncounted and approximations in the kits prices \textbf{($C_1$)}. 

Finally, the global cost difference is less than 4\% (Table \ref{tab:Tours}), mainly due to those "Uncounted" features of the GPON cost function that were not easily reproducible in NTS. 

\subsection{Discussion on Rouen }
The results for Rouen were analysed in the same detailed way and the differences have similar  causes and trends than in Tours (Table \ref{tab:Rouen}): (i) the global cost difference is less than 4 \%, mainly due to "Uncounted" (ii) the difference regarding the cost of the cables is less than 1 \% and (iii) the error on the variable part of the total costs is less than 8\%, which is precise enough both to compare different scenarios and to fix a budget.

\begin{table}
\renewcommand{\arraystretch}{1.3}
\caption{Total costs in Rouen: difference $\bigtriangleup$ between NTS and GPON. Policy 2 is used for Distribution.}
\label{tab:Rouen}
\centering
\begin{tabular}{cccc||c}
\hline 
 & Fixed & Variable & Total & Cables \\ 
\hline 
Transport & 0 \% & -22 \% & -8 \% & 1 \% \\ 
Distribution& 0 \% & 5 \% & -5 \%& 1 \% \\ 
Junction & -2 \% & 0 \% & -2 \% & 0 \% \\ 
Total & -2 \% & -8 \% & -4 \% & -1 \% \\ 
\hline 
\end{tabular} 
\end{table}

\section{Conclusion}
\label{sec:conclusion}
The theoretical results developed in previous works were implemented in two user-friendly prototypes for cost estimation of fixed access network deployment on any urban area with very short computing time. The idea is to let the user try several architectures and scenarios in order to compare them and determine the best one according to its criteria. On a specific use case (deploying optical fibre) the costs obtained are very close to those computed by an Orange optimization tool based on an entirely different approach. The NTS methodology based on statistical modelling proved to be quite efficient even on the small areas considered in the comparison. The tools are modular and will be enriched by non-isotropic models for streets (Manhattan) and extended to rural areas. 

The range of applications of OSMMiner/NTSTool prototypes is quite large since they make it possible to consider any architecture and scenario on any urban or rural area. We are currently considering  using their results on  the fibre networks linking mobile base stations to antennas, as well as on electrical networks. Another possible application is to evaluate the distance to drive in order to maintain all the mobile sites in a country, depending on the number of maintenance centres: this could allow us to better challenge the costs declared by the sub-contractors.

\section*{Appendix : Retreiving model parameters from reality}
\label{app:def}
A two levels network with a number $N_H$ of high nodes HL and $N_L$ of low nodes LL is to be deployed on a territory $T$ of area $A$ (km$^2$). Let us assume that the fitting process identified a PVT model of intensity $\gamma$ (km$^{-2}$) for the streets and recall that the corresponding morphological vector is 
$$\mathcal{M}_{theo}^{PVT}(\gamma)=(2 \gamma,3 \gamma,\gamma,\tau= 2 \sqrt{\gamma}),$$ 
where $\tau$ (km$^{-1}$) is the average length of streets par unit area. 

The averaged area of a HL serving zone is then $A /N_H$ since everything is assumed homogeneous. It contains a mean length of streets $L_s= \tau A /N_H$ (km) and is assimilated to the area of the typical "Poisson-Cox-Vorono\"i" cell centred on HL. This typical cell also contains by construction a unique HL thus giving $\lambda_{HL}=1/L_s$ (km$^{-1}$) for the linear intensity of the random process that models HL locations on the streets. A key parameter is the dimensionless ratio $\kappa= \tau / \lambda_{HL} =4 \gamma A /N_H.$
A high (resp. low) value of $\kappa$ describes a typical cell with a dense (resp. sparse) street system.  

Using scaling arguments, Palm theory and statistical analysis of extensive simulation data, it is shown that for $\kappa \ge 1$ the probability distribution of length $\ell$ (km) writes: 
$$dis_{LH}^{PVT} (\gamma,\lambda_{HL},\ell)=\tau W_t(a(\kappa), b(\kappa),c(\kappa),\tau\ell) $$
where $a, b, c $ are known polynomial functions and $W_t$ is the truncated Weibull probability distribution 
$$W_t(a,b,c,x)=\frac{a}{b} C^a (\frac{x}{b}+C)^{a-1} e^{(\frac{x}{b}+C)^a}$$
with $C=(bc/a)^{1/(a-1)}$. Explicit formulas exist for the cumulative function, quantile and average of $W_t$ distribution. Theoretical considerations provide formulas for $\kappa <1$.  Similar expressions hold for the others PLT and PDT models, using the associated expression for $\tau$. 
\bigskip

\textbf{Acknowledgments} The authors would like to thank M Jean-Philippe Lanquetin  and  M Patrick Boss\'e (Orange engineers), as well as  M Yu Hao, Ms Wenting Bao and Ms Lena Schmidt (Orange interns) who participated in the development of the two prototypes.

\end{document}